\documentclass[12pt,preprint]{emulateapj}

\usepackage{graphicx}
\usepackage{color}

\newcommand{\beq}{\begin{equation}}
\newcommand{\eeq}{\end{equation}}
\newcommand{\beqa}{\begin{eqnarray}}
\newcommand{\eeqa}{\end{eqnarray}}

\begin{document}

\title{ Looking for the gluon condensation signature in proton using the Earth limb gamma-ray spectra}

\author{ Lei Feng$^1$, Jianhong Ruan$^2$, Fan Wang$^3$ and  Wei Zhu$^2$\altaffilmark{*}}

\affiliation{ $^1$ Key Laboratory of Dark Matter and Space Astronomy, \\Purple Mountain Observatory,Chinese Academy of Sciences, Nanjing 210008, China}
\affiliation{ $^2$Department of Physics, East China Normal University, Shanghai 200241, China\\ }
\affiliation{$^3$Department of Physics, Nanjing University, Nanjing,210093, China} 
\altaffiltext{*}{ Corresponding author: wzhu@phy.ecnu.edu.cn}

\begin{abstract}
A new type of gamma ray spectrum is predicted in a general hadronic
framework by taking into account the gluon condensation effects in
proton. The result presents the power-law with a sharp break in the
gamma ray spectra at the TeV-band. We suggest to probe this
GC-signature in the Earth limb gamma-ray spectra using the DArk
Matter Particle Explorer and the CALorimetric Electron Telescope on
orbit.

\end{abstract}

\keywords{Gluon condensation: Broken power-law: Earth limb gamma-ray spectra:  DAMPE:  CALET}

\setlength{\parindent}{.25in}

\section{Introduction}
 A QCD study predicts
that gluons in proton may converge at a critical momentum
$(x_c,\underline{k}_c)$ (Fig.1), where $x$ is the fraction of
longitudinal momentum and $\underline{k}$ is the transverse momentum
carried by gluons (Zhu et al. 2008; 2016; 2017). This is the gluon condensation (GC). The GC
should induce significant effects in the proton collision processes,
provided the collision energy is higher than the GC-threshold
$E_{p-p(A)}^{GC}$.

\begin{figure}[htb]
    \centering {
        \includegraphics[width=0.96\columnwidth,angle=0]{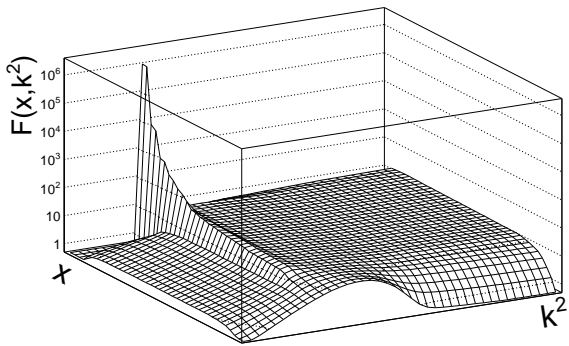}
    } \caption{\label{fig:fig1} A solution of the
        Zhu-Shen-Ruan equation (Zhu et al. 2016; 2017), which shows the evolution of
        transverse momentum dependent distribution of gluons in proton
        from a saturation input to the GC.}
\end{figure}

The energy of proton accelerated inside some sources, such as
supernova remnants (SNRs), active galactic nuclei (AGN) or pulsars,
could reach a very high level. Observations of such galactic
diffuse-ray emission (GDE) have provided valuable information about
cosmic rays. The power-law is a general form of the cosmic ray
spectra at high energy. It is described by a straight line of the
energy spectra with a fixed index in a log-log representation. This
line may span over more than one order of magnitude. The power-law
has a QCD explanation (Wong et al. 2015). On the other hand, break
in power-rule also has been observed and discussed in many works
(Yalcin et al. 2018). The broken power-law relates to the extra
sources of cosmic rays or even a new effect.
 The relativistic
protons with energy exceed $E_{p-p(A)}^{GC}$ collide with proton or
nuclei and produce vast number of photons through $p+p(A)\rightarrow
\pi^0\rightarrow 2\gamma$. One can image that these excessive
photons break the smooth $\gamma$-background at the break energy
$E_{\pi}^{GC}$.

In this report, we try to find an effective way to identify the
GC-signature in cosmic $\gamma$-rays. We noticed that there are
different gamma radiation mechanisms. Usually even one $\gamma$-ray
spectrum may lead to the debate of leptonic or hadronic
explanations. Therefore, we ask what is the GC-characteristic
spectrum? Can it distinguish the GC-effects from other phenomena?
For this sake, we give a brief review about the GC-effects in Sec.2.
Then we study the GC-effects in a general hadronic framework at
Sec.3. We derived an analytic solution of $\gamma$-ray spectra with
the GC-effects. The results present clearly a sharp broken
power-law, which is different from all other smoothly $\gamma$-ray
spectra.

However, a serious of uncertainties in hadronic process may hinder
our judgement for the GC-signature. (i) We cannot be sure the
acceleration mechanism and the primary proton spectra in the
different environments; (ii) What is the target nucleus and its
abundance? (iii) The complex interactions of photons with
interstellar medium, which include electron-positron
pair-production, ionization, diffusion and Compton scattering, may
influence the resulting spectra. Interestingly, the sharp broken
power-law has been recorded by the gamma-spectra in SRNs
(Archambault 2017; Condon et al. 2017), AGN (Zaborov et al. 2016) and
pulsars (Ackermann 2013). However, it is difficult to discriminate
whether they origin from the GC-effects or are produced by the
extragalactic background light (EBL) near the source (Abdalla et al.
2017). We will discuss them elsewhere.

We noticed that the above weakness can be skillfully complemented in
the Earth limb gamma-ray spectra. Such method was proposed in Ref.
(Thompson et al. 1981; Petry 2005; Abdo et al. 2009) and has been
used to probe the dark mater signal at $GeV$-band. We find that the
uncertainties in searching for the GC-signature may greatly reduced
if using the Earth limb observation. Based on the above discussions,
we propose it is possible to look for the GC-signature in the DArk
Matter Particle Explorer (DAMPE) (Ambrosi et al. 2017) and the
CALorimetric Electron Telescope (CALET) (Torii et al. 2015). We will
detail them at Sec. 4 and then give a summary at Sec. 5.

\section{A brief review about the gluon condensation}
 Gluons are Boson. The gluon condensation is an interesting
subject for long time. QCD analysis shows that the evolution
equation of gluons becomes nonlinear due to the correlations among
initial gluons at high energy and it results a balance between gluon
splitting and fusion, which is called as the color glass condensate
(CGC), where ``condensate" implies that the maximum occupation
number of gluons is $\sim 1/\alpha_s> 1$, although it lacks a
characteristic sharp peak in the momentum distribution (Jalilian-Marian et al. 1997; Jalilian-Marian et al 1997; Weigert 2002).

An advance of the QCD evolution equations is that a continuing
evolution of the CGC solution leads to the chaos solution (Zhu et
al. 2008; 2016; 2017). Most surprisingly, the dramatic chaotic
oscillations produce strong shadowing and antishadowing effects,
they converge gluons to a state with a critical momentum
$(x_c,\underline{k}_c)$. According to QCD, the number of secondary
particles (they are mostly pions) at the high energy $p-p(A)$
collisions relates to the number of gluons, which participate in the
multi-interactions. Pions will rapidly grow when a lot of gluons
enter the interaction range due to the GC effects. Without concrete
calculations, one can image that this will form an excess phenomenon
in the $\gamma$-ray spectra.

The quantitative calculations of the pion-distributions at the
$p-p(A)$ collisions are very complicated due to the nonperturbative
hadronization. For the simplification, we consider only pions as the
secondary particles since the multiplicities of other particles at
high energy collisions are much smaller than that of pions. Usually,
these pions have small kinetic energy (or low momentum) at the
center-of-mass (C.M.) system and form the central region in the
rapidity distribution. The maximum number of pions  $N_{\pi}$ at a
given interaction energy corresponds to a case, that all available
kinetic energy of the collide at the C.M. system are almost used to
create pions. It leads to $N_{\pi}\sim \sqrt {s}$. However, the data
show that $N_{\pi}\sim \ln s$ or $\ln s^2$ (Anisovch et al. 1985). A
possible reason is that the limited available number of gluons
restricts the increase of secondaries (Wang et al. 1995). We assume
that a large number of gluons at the central region due to the
GC-effects create the maximum number $N_{\pi}$ of pions. We
emphasize that this assumption is a simplification method rather
than a necessary GC-condition. In fact, we will show that it may
simplify the calculation but does not essentially£©change the
GC-characteristic signature essentially. Using general relativistic
invariant and energy conservation, we have

$$(2m_p^2+2E_{p-p(A)}m_p)^{1/2}=E^*_{p1}+E^*_{p2}+N_{\pi}m_{\pi}, \eqno(2.1)$$
$$E_{p-p(A)}+m_p=m_p\gamma_1+m_p\gamma_2+N_{\pi}m_{\pi}\gamma, \eqno(2.2)$$where
$E^*_{p_i}$ is the energy of leading proton at the C.M. system,
$\gamma_i$ are the corresponding Lorentz factors. Using the
inelasticity $K$ (Gaisser 1990), we set

$$E^*_{p1}+E^*_{p2}=(\frac{1}{K}-1)N_{\pi}m_{\pi},  \eqno(2.3)$$and

$$m_p\gamma_1+m_p\gamma_2=(\frac{1}{K}-1) N_{\pi}m_{\pi}\gamma. \eqno(2.4)$$
One can easily get the solutions $N_{\pi}(E_{p-p(A)},E_{\pi})$ for
the $p-p(A)$ collisions

$$\ln N_{\pi}=0.5\ln E_{p-p(A)}+a, ~~\ln N_{\pi}=\ln E_{\pi}+b,\eqno(2.5)$$  where $E_{\pi}
\in [E_{\pi}^{GC},E_{\pi}^{max}]$.  The parameters

$$a\equiv 0.5\ln (2m_p)-\ln m_{\pi}+\ln K, \eqno(2.6)$$ and

$$b\equiv \ln (2m_p)-2\ln m_{\pi}+\ln K. \eqno(2.7)$$ Equation (2.5) gives
the one-to-one relation among $N_{\pi}$, $E_{p-p(A)}$ and
$E_{\pi}^{GC}$, which leads to a GC-characteristic spectrum.

\section{The GC effects in gamma ray spectra}

 Imaging a high energy proton collides with a proton or a
nucleus, we have $p+p(A)\rightarrow \pi^{\pm,0}+others$ and
following $\pi^0\rightarrow 2\gamma$. The corresponding gamma flux
in a general hadronic framework reads

$$\Phi_{\gamma}(E_{\gamma})=\Phi^0_{\gamma}(E_{\gamma})+
\Phi^{GC}_{\gamma}(E_{\gamma}),\eqno(3.1)$$$\Phi^0_{\gamma}(E_{\gamma})$
is the background contributions and

$$\Phi^{GC}_{\gamma}(E_{\gamma})=C_{p-p(A)}\left(\frac{E_{\gamma}}{E_0}\right)^{-\beta_{\gamma}}
\int_{E_{\pi}^{min}}^{E_{\pi}^{max}}dE_{\pi}
\left(\frac{E_{p-p(A)}}{E_{p-p(A)}^{GC}}\right)^{-\beta_p}$$\\
$$\times{N_{\pi}(E_{p-p(A)},E_{\pi})
\frac{d\omega_{\pi-\gamma}(E_{\pi},E_{\gamma})}{dE_{\gamma}}},
\eqno(3.2)$$
where index $\beta_{\gamma}$ and $\beta_p$ denote the
propagating loss of gamma rays and the acceleration mechanism of
protons respectively; $C_{p-p(A)}$ incorporates the kinematic factor
with the flux dimension and the percentage of $\pi^0\rightarrow
2\gamma$. The normalized spectrum for $\pi^0\rightarrow 2\gamma$ is

$$\frac{d\omega_{\pi-\gamma}(E_{\pi},E_{\gamma})}{dE_{\gamma}}
=\frac{2}{\beta_{\pi}
    E_{\pi}}H[E_{\gamma};\frac{1}{2}E_{\pi}(1-\beta_{\pi}),
\frac{1}{2}E_{\pi}(1+\beta_{\pi})], \eqno(3.3)$$ $H(x;a,b)=1$ if
$a\leq x\leq b$, and $H(x;a,b)=0$ otherwise. Inserting Eq. (2.5) and
(3.3) into Eq. (3.2), we have

$$\Phi^{GC}_{\gamma}(E_{\gamma})=C_{p-p(A)}\left(\frac{E_{\gamma}}{E_{\pi}^{GC}}\right)^{-\beta_{\gamma}}
\int_{E_{\pi}^{GC}~or~E_{\gamma}}^{E_{\pi}^{GC,max}}dE_{\pi}$$
$$\times{\left(\frac{E_{p-p(A)}}{E_{p-p(A)}^{GC}}\right)^{-\beta_p}N_{\pi}(E_{p-p(A)},E_{\pi})
\frac{2}{\beta_{\pi}E_{\pi}}}, \eqno(3.4)$$where the low-limit of the
integration takes $E_{\pi}^{GC}$ (or $E_{\gamma}$) if
$E_{\gamma}\leq E_{\pi}^{GC}$ (or if $E_{\gamma}> E_{\pi}^{GC}$). In
consequence,

$$E_{\gamma}^2\Phi^{GC}_{\gamma}(E_{\gamma})$$
$$=\left\{
\begin{array}{ll}
\frac{2C}{2\beta_p-1}e^b(E_{\pi}^{GC})^3\left(\frac{E_{\gamma}}{E_{\pi}^{GC}}\right)^{-\beta_{\gamma}+2} & {\rm if~}E_{\gamma}\leq E_{\pi}^{GC}\\\\
\frac{2C}{2\beta_p-1}e^b(E_{\pi}^{GC})^3\left(\frac{E_{\gamma}}{E_{\pi}^{GC}}\right)^{-\beta_{\gamma}-2\beta_p+3}
& {\rm if~} E_{\gamma}>E_{\pi}^{GC}
\end{array} \right. .\eqno(3.5)$$It is the power-law with a sharp break.
The break energy $E_{\gamma}=E_{\pi}^{GC}$ is a direct result of the
gluon distribution in Fig.1, where a sharp peak divides the spectrum
into two parts. A pure power-law at $E_{\gamma}<E_{\pi}^{GC}$
origins from a fixed lower limit $\sim E_{\pi}^{GC}$ of integral in
Eq. (3.4) and it is irrelevant to the concrete form of Eq. (2.5).
The above two universal behaviors of $\Phi_{\gamma}^{GC}$ are
directly arisen from the GC-effects and they are different from all
other well known smoothly radiation spectra. We regard them as the
GC-character. The second power-law at $E_{\gamma}>E_{\pi}^{GC}$ is a
simplified result in Eqs. (2.1) and (2.2), where all available
kinetic energies in the central region are used to create pions. We
emphasize that any deformations from this power-law at
$E_{\gamma}>E_{\pi}^{GC}$ are allowed if our simplification is
modified. Nevertheless, it does not change the above mentioned
GC-character. One can compare Eq. (3.5) with the experimental data
to check the validity of the simplification.

We compare the gamma-ray spectra with and without the GC-effects in
a same hadronic framework. The second case was taken as a first
evidence of hadronic component in cosmic ray spectra (Ackermann et
al. 2013; Dermer et al.  2013). The $p-p(A)$ collisions create
$\pi^0$. After a mean lifetime of $8\times 10^{-17}s$, $\pi^0$
decays into two gamma-photons with a given energy of $m_{\pi}=67.5~
MeV$ in the rest frame of the pion. This energy distribution will be
substantially Doppler shifted with respect to the Earth rest frame
due to the large kinetic energies of the $\pi^0$ mesons
(Fig.2(b,e,g)). In general case, $\pi$-spectra $N_{\pi}(E_{p-p(A)},
E_{\pi})$ at p-p(A) collisions in Eq.(3.2) are complicated function
because they contain unknown hadronization dynamics in
multiproduction. In the case without the GC-effects (Dermer et al.
2013), the authors used a two-body model at the low energy
approximation: $p+p\rightarrow \Delta$, $\Delta\rightarrow
p+p+\pi^0$ and $\pi^0\rightarrow 2\gamma$. The resulting $N_{\pi}$
is a smooth function of $E_{\pi}$ defined in the range
$[E_{\pi}^{min},E_{\pi}^{max}]$ (Fig.2(a)). After integral in
Eq.(3.2), one can get a smooth excess curve, which peaks at $\sim
1~GeV$ in $E_{\gamma}^2\Phi\sim E_{\gamma}$ plot (see Fig.2(c)).
Note that the factor $E_{\gamma}^{-\beta_{\gamma}}$ is neglected in this example. The
results are demonstrated by the W44- and IC443-gamma spectra
(Ackermann et al. 2013).

We now consider the GC-effects (Fig.2(d-h)). A main difference is
that $N_{\pi}$ takes Eq. (2.5), which is defined in the range
$E_{\pi}\in [E_{\pi}^{GC}, E_{\pi}^{max}]$. The conditions
$E_{\gamma}\leq E_{\pi}^{GC}$ and $E_{\pi}\geq E_{\pi}^{GC}$ divide
the integral into two parts. Thus the power-law is broken at
$E_{\pi}^{GC}$ (see dashed lines (or solid lines) in Fig.2(h) if
without (or with) the corrections of $E_{\gamma}^{-\beta_{\gamma}}$
and $E_p^{-\beta_p}$).

Equation (3.5) is the solution of analytic form for the
GC-characteristic spectra rather than a numerical simulation. Their
parameters have the definite physical meaning. Therefore, a
deviation from the model Eqs. (2.1) and (2.2) may destroy the line
at $E_{\gamma}>E_{\pi}^{GC}$ in the log-log representation. While
the pure power-law at $E_{\gamma}<E_{\pi}^{GC}$ in Eq.(3.5) is
irrelevant to Eq.(2.5), unless the power form
$E_{\gamma}^{-\beta_{\gamma}}$ is broken. Thus, we can modify our
assumption in Eq.(2.5) according to the derivations of the predicted
broken power-law Eq. (3.5).

\begin{figure}[htb]
    \centering {
        \includegraphics[width=0.96\columnwidth,angle=0]{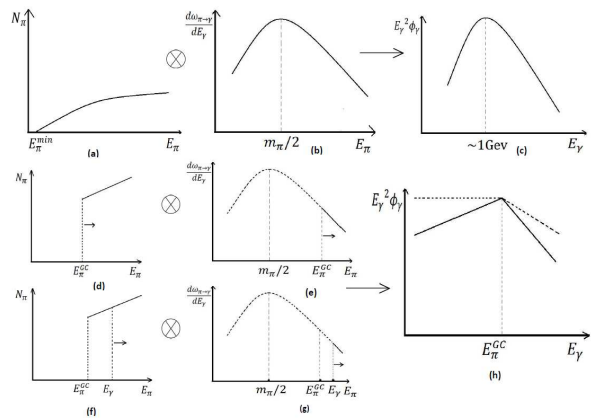}
    } \caption{\label{fig:fig2}Schematic diagrams for $\gamma$-ray spectra in the hadronic model.
        (a-c) without the GC-effects; (d-h) with the GC-effects.}
\end{figure}

\section{The Earth limb photon spectra at GeV-TeV band}
The observation of the Earth limb gamma-ray spectrum is shown in
Fig.3. Cosmic rays (mainly protons) entering the atmosphere produce
a lot of secondary particles including photons due to the $p-air$
collisions. A considerable amount of these $\gamma$-rays spreads
toward the dense atmosphere and form high energy showers, which may
mix with the $\gamma$-rays originating from the $p-air$
interactions. However, a part of the oblique incidence photons
propagate toward the thin atmosphere and we can neglect the
contributions of the air-shower. These $\gamma$-spectrum depends on
the $p-air$ cross section and the spectrum of the incident protons.
One can measure this so called the Earth limb $\gamma$-spectrum by
changing the orbit and direction of the detector.

\begin{figure}[htb]
    \centering {
        \includegraphics[width=0.8\columnwidth,angle=0]{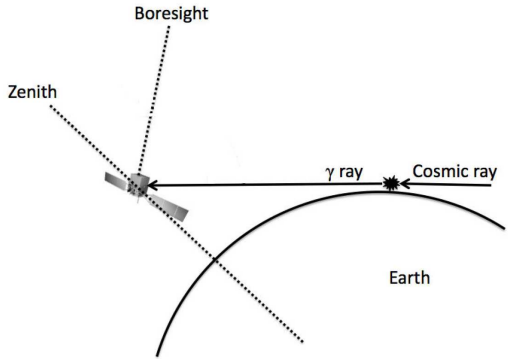}
    } \caption{\label{fig:fig3}  Schematic of the Earth Limb gamma ray production by
        cosmic rays from the Earth atmosphere (Cited from
        https://fermi-hero.readthedocs.io/en/latest/$\_$images/earth$\_$limb$\_$gammas.png.).}
\end{figure}

In contrast to the GDE, the contribution from the inverse Compton
scattering of cosmic ray electrons to the Earth limb emission is
negligible, because the atmosphere lacks dense soft photons and
strong electron flux. It avoids the argument between leptonic and
hadronic models. The spectrum of incident protons near the
atmosphere is well known and spectrum index is $\beta_p\simeq2.75$,
while the target nuclei (mainly nitrogen and oxygen) and their
abundances are completely fixed. Besides, the observation at the top
of the atmosphere may reduce the influence of electromagnetic
shower.

In order to illustrate the GC-spectra in the p-N(O) collisions, we
need to know the value of $E_{\pi}^{GC}$, which is target-dependent.
Because the nonlinear term of the QCD evolution equation should be
re-scaled by $A^{1/3}$, $E_{\pi}^{GC}$ decreases with increasing A
(Zhu et al.2016). Therefore, $E_{p-A}^{GC}<E_{p-p}^{GC}$ is
expected. In this report we take the following estimation. Since the
$Pb-Pb$ collisions have not found the GC-signature at LHC till
$\sqrt{s_{Pb-Pb}}\sim 8~TeV$, we suggest that
$E_{p-Pb}^{GC}>3.4\times 10^7~GeV$ since
$\sqrt{s_{p-Pb}}=\sqrt{2m_pE_{p-Pb}}$, i.e.,
$E_{\pi}^{GC}(p-Pb)>600~GeV$ according to Eq. (2.5), where we take
$K=0.5$ (Gaisser 1990).

The cross section $\sigma_{p-air}$ was estimated using the
distribution of the shower maximum slant depth $X_{max}$, where the
interaction length and consequently the $\sigma_{p-air}$ is related
to the exponential tail of the $X_{max}$ distribution('s exponential
tail) in the cascade model. The result of Telescope Array has not
found a big increment effect in $\sigma_{p-air}$ till
$\sqrt{s}=95~TeV$ (Hanlon et al. 2017; Abbasi et al. 2018). However,
the cascade method has the fundamental limitation. Part of important
parameters are uncertain (Gaisser 1990). Besides, the GC-effects may
disappear quickly at the beginning of collisions due to the energy
loss of the leading proton. Therefore, we do not regard the
Telescope Array data as a restriction to the value of $E_{\pi}^{GC}$
in this work.

The bright gamma-ray emission from the Earth limb was observed by
the SAS-2 (Thompson et al. 1981), EGRET (Petry 2005) and Fermi-LAT
(Ackermann et al. 2014) instruments. The Earth¡¯s limb $\gamma$-ray
energy spectra have a power-law $\sim E_{\gamma}^{-2.75}$ and their
measurements reach $E_{\gamma}\sim 400~GeV$ (Ackermann et al. 2014).
If this law is valid at even higher energy, one can estimate that
the $\gamma$-flux will be reduced by 3 order at $E_{\gamma}=4~TeV$
(or 4 order at $E_{\gamma}=10~TeV)$. On the other hand, as we have
estimated that the inelastic cross section of $p-p$ collision
increases by 4 order due to the GC-effects (Zhu et al. 2017).
Therefore, the observation of the GC-signature in the Earth limb
$\gamma$-energy spectrum is possible. We noticed that DAMPE and
CALET have higher resolution and adjustable observation method. The
energy range for DAMPE is $5~ GeV\sim 10~ TeV$ which is similar to
that of CALET. They are suitable for probing the GC-signature in the
Earth limb gamma-ray spectrum. Fig. 4 gives a schematic diagram for
the contributions of the GC-effects to
$\Phi^{GC}_{\gamma}(E_{\gamma})$, where the resulting curve is not
scaled. A sharp peak originating from the GC-effects in the Earth
limb $\gamma$-energy spectrum is different from all other smoothly
excess and we can easily recognize it.

\begin{figure}[htb]
	\centering {
		\includegraphics[width=0.90\columnwidth,angle=0]{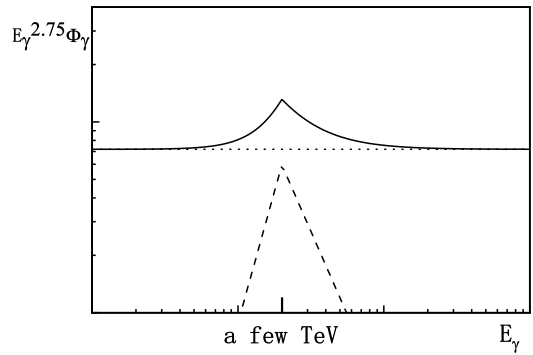}
	} \caption{\label{fig:fig4}  The GC-signature in the Earth limb gamma-ray
		spectra, which is not scaled. $\beta_{\gamma}=0$ and $\beta_p=2.75$.
		The resulting spectrum presents a GC-characteristic sharp peak. Dotted line is
		background $\sim E^{-2.75}_{\gamma}$}
\end{figure}

 The gamma ray signal of the dark matter (DM) aroused great interest.
If the DM annihilates to gamma-rays directly, one can get a
gamma-ray line signature. While the DM produces leptons and quarks,
the gamma spectrum as a final product presents an "excess'' or a
smooth broken power-law. These two forms are obviously different
from the GC effects-a sharp broken power. Besides, it is impossible
that the DM generates a new Earth limb gamma-signal through dark
matter annihilation in the atmosphere, since the cross section and
the DM density in the atmosphere region are all very small and the
corresponding gamma flux is negligible. Therefore, we can
distinguish the GC-signature from the DM-signals in the Earth limb
gamma-ray spectra.

\section{Summary}
A research of QCD evolution dynamics shows that the gluons in
proton may converge to a state with a critical momentum at a high
energy range. The GC-effects should produce a lot of extra secondary
pions at the central region of the rapidity distributions. Without
concrete calculations, one can image that it will induce gamma-ray
excess if the interaction energy of the $p-p(A)$ collisions larger
than a critic scale. A quantitative analysis presents the
GC-characteristic gamma-spectrum, which is a sharp broken power-law.
We suggest the observation of Earth limb gamma-ray spectra at the
GeV-TeV-band using the DAMPE and CALET installations on orbit to
probe the GC-signature.

{\it Acknowledgments:} This work is supported by the National
Natural Science of China (No.11851303). F. L. acknowledges partial
support from the National Key Research and Development
Program of China (No. 2016YFA0400200), the National Natural Science of China (No. 11773075)
and the Youth Innovation Promotion
Association of Chinese Academy of Sciences\\       (No. 2016288).

\end{document}